\documentstyle[aps,prb,twocolumn,eqsecnum,epsfig,floats]{revtex}

\begin{document}

\draft \preprint{}

\twocolumn[\hsize\textwidth\columnwidth\hsize\csname
@twocolumnfalse\endcsname

\title{Creation of nonlocal spin-entangled electrons via Andreev
tunneling,\\ Coulomb blockade and resonant transport}
\author{Patrik Recher and Daniel Loss}

\address{
Department of Physics and Astronomy, University of Basel,\\
Klingelbergstrasse 82, CH-4056 Basel, Switzerland}

\date{\today}

\maketitle
\begin{abstract}
We discuss several scenarios for the creation of nonlocal
spin-entangled electrons which provide a source of electronic
Einstein-Podolsky-Rosen (EPR) pairs. Such EPR pairs can be used to
test nonlocality of electrons in solid state systems, and they
form the basic resources for quantum information processing. The
central idea is to exploit the spin correlations naturally present
in superconductors in form of Cooper pairs possessing spin-singlet
wavefunctions. We show that nonlocal spin-entanglement in form of
an effective Heisenberg spin interaction is induced between
electron spins residing on two  quantum dots with no direct
coupling between them but each of them being tunnel-coupled to the
same superconductor. We then discuss a nonequilibrium setup with
an applied bias where mobile and nonlocal spin-entanglement can
be created by coherent injection of two electrons in a pair
(Andreev) tunneling process into two spatially separated quantum
dots and subsequently into two Fermi-liquid leads. The current for
injecting two spin-entangled electrons into different leads shows
a resonance and allows the injection of electrons at the same
orbital energy, which is a crucial requirement for the detection
of spin-entanglement via the current noise. On the other hand,
tunneling via the same dot into the same lead is suppressed by the
Coulomb blockade effect of the quantum dots. We discuss
Aharonov-Bohm oscillations in the current and show that they
contain $h/e$ and $h/2e$ periods, which provides an experimental
means to test the nonlocality of the entangled pair. Finally we
discuss a structure consisting of a superconductor weakly coupled
to two separate one-dimensional leads with Luttinger liquid
properties. We show that strong correlations again suppress the
coherent subsequent tunneling of two electrons into the same lead,
thus generating again  nonlocal spin-entangled electrons in the
Luttinger liquid leads.
\end{abstract}
\vskip2pc] \narrowtext

\section{Introduction}
The electron has charge as well as spin. While the control of
charge is well established in electronic systems and devices,
 the spin combined with transport only recently attracted  interest,
 both from a fundamental
 point of view and also for applications in electronics \cite{Prinz,Prinz2}.
 The idea of using the spin and its coherence for electronic devices
 has received strong experimental
 support \cite{Kikkawa1,Kikkawa2,Awschalom,Fiederling,Ohno} showing
unusually long
 spin dephasing times \cite{Kikkawa1,Kikkawa2,Awschalom}
 in semiconductors (approaching microseconds),
 the injection of spin-polarized currents from a magnetic to a nonmagnetic
 semiconductor\cite{Fiederling,Ohno},
 as well as phase-coherent spin transport
 over distances of up to $100\:\mu{\rm m}$
\cite{Kikkawa1,Kikkawa2,Awschalom}.
 Besides the improvement of conventional devices \cite{Prinz,Prinz2}
 due to spin, e.g. in magnetic read out heads, non-volatile memories etc.,
 electron spins in quantum confined nanostructures like semiconductor
 quantum dots, atoms, or molecules
 can be used as a quantum bit (qubit)~\cite{Loss97}
 for quantum computing~\cite{Steane,MMM2000} and quantum
 communication~\cite{MMM2000,Bennett00},
 where a radically new design of the necessary hardware
 is required.  In this article we focus on the creation of spin-entangled
 electrons--electronic Einstein-Podolsky-Rosen (EPR) pairs
\cite{Einstein}--exploiting the spin-correlations naturally
present in s-wave superconductors, where the electrons are
correlated as Cooper
 pairs with spin-singlet wavefunctions.
We note that such EPR pairs represent
 the basic resources for quantum
 communication~\cite{Bennett00} schemes like dense
 coding, teleportation, or, more fundamentally,
 for testing nonlocality via Bell inequalities\cite{Bell}, which would
 be particularly interesting to implement for massive particles such as
 electrons
 in a condensed matter system~\cite{MMM2000,BLS}.\\

 We describe in the following a few proposals for creating nonlocal spin-entanglement
between two electrons. We consider first two  quantum dots  which
are coupled to the same superconductors (but not among
themselves). In {\em equilibrium}, this coupling then induces an
effective tunable Heisenberg interaction \cite{CBL} among the two
electrons on the dots which could be used to implement two-qubit
gates \cite{QCReview}. We then consider a {\em nonequilibrium}
situation where now Cooper pairs can be transported by means of an
Andreev (pair tunneling) process from the superconductor via two
quantum dots into {\em different } normal Fermi liquid leads by
applying a bias between the superconductor and the leads
\cite{RSL}. The quantum dots in the Coulomb blockade regime are
needed to mediate the necessary interaction between the two
tunneling electrons so that they preferably tunnel to different
dots and then subsequently to different outgoing leads, thereby
maintaining their spin-singlet state. Such a setup then works as
an entangler for electron spins, satisfying all necessary
requirements, within a parameter regime of experimental
accessebility, to detect the spin-entanglement in the leads via
the current noise \cite{BLS}. We finally discuss a further
realization for an entangler where the necessary interaction to
separate the two electrons is provided by strong correlations in
one-dimensional leads (such as nanotubes) with Luttinger liquid
properties.

 We refer  to
related work \cite{Lesovik} which makes also use of Andreev
tunneling but with a transparent superconductor/normal interface.
The electrons move from the superconductor into a normal
fork-shaped wire without Coulomb blockade behavior. The electrons
are separated via energy filters so that the electrons enter their
corresponding leads at different orbital energies. However, due to
the transparent interface the partners of different Cooper pairs
are not separated in time and space which is needed to identify
the spin-entangled partners. Subsequent work  in similar direction
uses multiterminal hybrid structures \cite{Falci,Melin}.

\section{Creation and Detection\\ of nonlocal\\
Spin-Entanglement in a Double Dot}

We consider a parallel double quantum dot structure. The quantum
dots contain one electron spin each and we assume that there is no
direct coupling between the dots but each dot is tunnel-coupled to
the same s-wave superconductors with a tunnel amplitude $\Gamma$
(see Fig.\ 1) on both sides of the dots. The two superconductors
are held at the same chemical potential.
\begin{figure}[h]
\centerline{\psfig{file=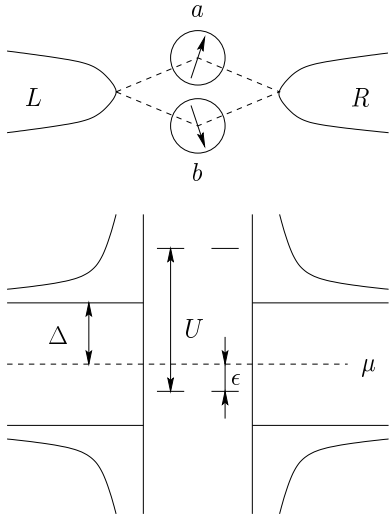,width=6cm}} \vspace{5mm}
\label{spintabl} \caption{Upper panel: Sketch of the
superconductor-double quantum
 dot-superconductor (S-DD-S) nanostructure. There is no direct coupling between
 the dots. Lower panel: Schematic
 representation of the quasiparticle energy spectrum in
the superconductors and
 the single-electron levels of the two quantum dots.}
\end{figure}
The s-wave superconductor favors an entangled singlet-state on the dots
(like in a Cooper pair)
 and further provides a mechanism for detecting the spin state via the
Josephson
current.
It turns out that in leading order $\propto \Gamma^4$ the spin coupling is
described by
 a Heisenberg  Hamiltonian \cite{CBL,CBL2} with an antiferromagnetic exchange coupling $J$
\begin{equation}
H_{\rm eff}
\approx J\,(1+\cos\varphi)\,
 \left({\bf S}_a\cdot{\bf S}_b-\frac{1}{4}\right) \;,
\end{equation}
where $J\approx 2\Gamma^2/\epsilon$,
 and the energy of the dot is $\epsilon$ below the chemical potentials of the superconductors.
Here, $\varphi$ is the average phase difference across the
superconductor--double-dot--superconductor (S-DD-S) junction.
We can modify the exchange coupling between the spins by tuning
the external
control parameters $\Gamma$ and $\varphi$ so the device can act as a two-qubit quantum gate
needed for quantum computing where the electron spin is the fundamental computational unit \cite{Loss97,QCReview}.
Furthermore, the entangled spin state on the dot
 can be probed if the superconducting leads are
 joined with one additional (ordinary) Josephson junction with
 coupling $J'$ and phase difference $\theta$
 into a SQUID-ring.
The supercurrent $I_S$ through this ring is given by~\cite{CBL}
\begin{equation}
\label{eqnIsIjDD}
I_S/I_J
= \left\{\begin{array}{ll}
 \sin(\theta-2\pi{f}) + (J'/J)\sin\theta\, ,
 &\mbox{singlet}, \\
 (J'/J)\sin\theta\, ,
 &\mbox{triplets},
 \end{array}\right.
\end{equation}
where $I_J = 2eJ/\hbar$ and $f=\Phi/\Phi_{0}$ with $\Phi$ being the magnetic flux threading the SQUID-ring and $\Phi_{0}=\hbar c/2e$ is the flux quantum.
Measurement of the spin- and flux-dependent critical current
 $I_c=\max_{\theta} \{|I_S|\}$
 probes the spin state of the double dot.
This is realized by biasing the system with a dc current $I$
 until a finite voltage $V$ appears for $|I|>I_c$~\cite{CBL}.

\section{Andreev Entangler}

We now consider a { \em nonequilibrium} situation where the superconductor is tunnel-coupled to two quantum dots in the Coulomb blockade regime which are further coupled to two normal Fermi liquid leads \cite{RSL}, see Fig.\ 2. Applying a bias between the superconductor and the leads than results in a stationary current of spin-entangled electrons via Andreev tunneling and resonant transport from the superconductor to the leads. The quantum dots are used to mediate the interaction necessary to separate the two spin-entangled electrons originating from a Cooper pair.
The amount of spin-entanglement in the outgoing current could be tested via noise measurements \cite{BLS}. We have shown that current-current correlations (noise) are enhanced if the injected electrons are singlets due to bunching behavior whereas the noise is suppressed in the case of spin triplets due to antibunching behavior. For such noise measurements, which are based on
two-particle interference effects, it is
absolutely crucial that both  electrons, coming from
different
leads, possess the {\em same} orbital energy.

\begin{figure}[h]
\centerline{\psfig{file=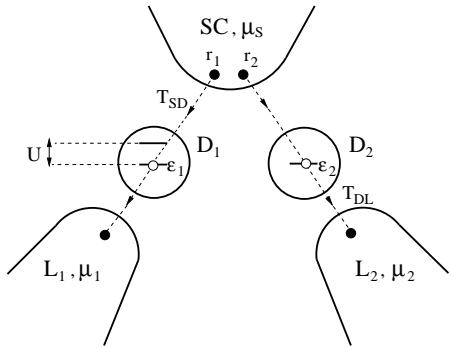,width=6cm}}
\vspace{5mm}
\label{spintabl}
\caption{The entangler setup: Two spin-entangled electrons forming a Cooper pair
can tunnel with amplitude $T_{SD}$ from points ${\bf r}_{1}$ and ${\bf
r}_{2}$ of the superconductor, SC, to two dots, $D_{1}$ and $D_{2}$, by means
of Andreev tunneling. The dots are tunnel-coupled to noninteracting normal leads ${\rm L_{1}}$
and ${\rm L_{2}}$, with tunneling amplitude $T_{DL}$.
The superconductor and leads are kept at chemical potentials $\mu_{S}$ and
$\mu_{l}$, resp.}
\end{figure}

\section{Qualitative Description of the Andreev Entangler}

We first provide a qualitative description of the entangler and
its principal mechanism based on Andreev processes and Coulomb
blockade effects and also specify the necessary parameter regime
for successful transport of the initial spin-entanglement of the
Cooper pairs to the outgoing leads. In subsequent sections we then
introduce the  Hamiltonian and calculate the stationary current
for two competing transport channels which is followed by a
discussion of the results.

We consider an s-wave superconductor
where the elec-\\trons form Cooper pairs  with singlet
spin-wavefunctions\\\cite{Schrieffer}. The superconductor, which is
held at the chemical potential $\mu_{S}$, is weakly coupled by
tunnel barriers to two separate quantum dots $D_{1}$ and $D_{2}$
which themselves are weakly coupled to Fermi liquid leads $L_{1}$
and $L_{2}$, resp., both held at the same chemical potential
$\mu_{1}=\mu_{2}=\mu_{l}$. The corresponding tunneling amplitudes between
superconductor and dots, and dot-leads, are denoted by $T_{SD}$
and $T_{DL}$, resp. which, for simplicity,  we assume to be equal
for both dots and leads.

By applying a bias voltage $\Delta\mu=\mu_{S}-\mu_{l}>0$ transport
of entangled electrons occurs from the superconductor via the dots
to the leads. In general, the tunnel-coupling of a superconductor
to a normal region allows for coherent transport of two electrons
of opposite spins due to Andreev tunneling\cite{Schrieffer}, while
single-electron tunneling is suppressed\cite{Glazman} in the
regime $\Delta>\Delta\mu,k_{B}T$, where $\Delta$ is the energy gap
in the superconductor and $T$ is the temperature. The gap $\Delta$
is the minimum energy to break up a Cooper püair into a
quasiparticle in the superconductor and an electron in the normal
region due to tunneling. According to the energy-time uncertainty
relation, $\hbar/\Delta$ then defines the time delay between the
two coherent tunneling steps in the Andreev process. In the
present setup, we envision a situation where the two electrons are
forced to tunnel coherently into {\em different} leads rather than
both into the same lead. This situation can be enforced in the
presence of two intermediate quantum dots which are assumed to be
in the Coulomb blockade regime\cite{Kouwenhoven} so that the state
with the two electrons being on the same quantum dot is strongly
suppressed, and thus the electrons will preferably tunnel into
separate dots and subsequently into separate leads (this will be
quantified in the following).

The chemical potentials $\epsilon_{1}$ and $\epsilon_{2}$
of the quantum dots can be tuned by external gate
voltages\cite{Kouwenhoven} such that
the coherent tunneling of two electrons into
different leads is at resonance if
$\epsilon_{1}+\epsilon_{2}=2\mu_{S}$, see Fig.\ 3.
\begin{figure}[h]
\centerline{\psfig{file=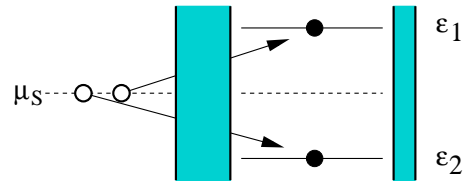,width=6cm}}
\vspace{5mm}
\label{Andreevdot2}
\caption{The energy situation of the superconductor with chemical potential $\mu_{S}$ and
the two dots 1,2 with chemical potentials $\epsilon_{1},\epsilon_{2}$. Transport of
the two members of a Cooper pair with energy $2\mu_{S}$ from the superconductor to {\em different}
outgoing leads with chemical potential $\mu_{l}<\epsilon_{l}$ (not drawn) is at
resonance if $\epsilon_{1}+\epsilon_{2}=2\mu_{S}$.}
\end{figure}
This current resonance condition reflects energy conservation in a
tunnel process of a Cooper pair with energy $2\mu_{S}$ from the
superconductor to the dots 1,2 (one electron on each dot) with
chemical potentials $\epsilon_{1},\epsilon_{2}$ and requires that
the resonant dot levels have to be adjusted such that one is above
$\mu_{S}$ and the other (by the same amount) below  $\mu_{S}$.
This is very similar to the more familiar picture of Andreev
reflection at a superconductor/normal interface. There an electron
on the normal side of the junction, and with energy $\epsilon$
above $\mu_{S}$, is back reflected as a hole with energy
$\epsilon$ below $\mu_{S}$ by the simultaneous creation of a
Cooper pair in the superconductor. In that sense the empty dot
level below $\mu_{S}$ can be considered as  the hole and the empty
dot level above $\mu_{S}$ as  the empty electron state. In
contrast, we will see that the current for the coherent tunneling
of two electrons via the {\it same} dot into the {\it same} lead
is suppressed by the on-site Coulomb $U$ repulsion of a quantum
dot and/or by the superconducting gap $\Delta$.

Next, we introduce  the relevant parameters describing the proposed
device and specify
their regime of interest.
First we note that to
avoid unwanted correlations with electrons already on the quantum dots
 one could work
in the cotunneling regime\cite{Kouwenhoven} where the number of
electrons on the dots are fixed and the resonant levels
$\epsilon_{l}$, $l=1,2$ cannot be occupied. However, we prefer to
work at the particular resonance $\epsilon_{l}\simeq\mu_{S}$,
since then the total current and the desired suppression of
tunneling into the same lead is maximized. Also, the desired
injection of the two electrons into different leads but at the
{\it same} orbital energy is then achieved. In the resonant
regime, we can avoid unwanted correlations between tunneling of
subsequent Cooper pairs if we require that the dot-lead coupling
is much stronger than the superconductor-dot coupling, i.e.
$|T_{SD}|< |T_{DL}|$, so that electrons which enter the dots from
the superconductor will leave the quantum dots to the leads much
faster than new electrons can be provided from the superconductor.
In addition, a stationary occupation due to the coupling to the
leads is exponentially small if $\Delta\mu>k_{B}T$, $T$ being the
temperature and $k_{B}$ the Boltzmann constant. Thus in this
asymmetric barrier case, the resonant dot levels $\epsilon_{l}$
can be occupied only during a virtual process (see also Section \ref{T-matrix}).

Next, the quantum dots in the ground state are allowed to contain an arbitrary but even
number of electrons, $N_{D}=\rm{even}$, with total spin zero
(i.e. antiferromagnetic filling of the dots).
An odd number $N_D$ must be excluded since a simple
spin-flip on the quantum dot would be possible in the transport process
and as a result the desired entanglement would be lost.
Further, we have to make
sure that also spin flip processes of the following kind are excluded.
Consider
an electron that tunnels from the superconductor into a given dot. Now, it
is
possible in principle (e.g. in a sequential tunneling
process\cite{Kouwenhoven})
that another electron
with the opposite spin leaves the dot and tunnels into the lead, and, again,
the
desired entanglement would be lost. However, such spin flip processes will
be
excluded if the energy level spacing of the quantum dots, $\delta\epsilon$,
(assumed
to be similar for both dots) exceeds both, temperature $k_{B}T$ and bias
voltage
$\Delta\mu$.
A serious source of
entanglement-loss is given by electron hole-pair excitations  out of
the Fermi sea of the leads during the resonant tunneling events. Since then a simple spin flip on the dot would be possible due to the coupling to the leads.
However,  we showed \cite{RSL} that such
many-particle contributions can be suppressed if the resonance width
$\gamma_{l}=2\pi\nu_{l}|T_{DL}|^2$ is smaller than $\Delta\mu$
(for $\epsilon_{l}\simeq
\mu_{S}$), where $\nu_{l}$ is the density of states (DOS) per spin of
the leads at the chemical potential $\mu_{l}$.

To summarize, the regime of interest where the coherence of an initially entangled Cooper pair (spin singlet) is preserved during the transport to the leads is given by
\begin{equation}
\label{regime}
\Delta, U,
\delta\epsilon>\Delta\mu>\gamma_{l},  k_{B}T,\quad
{\rm and}\quad \gamma_{l}>\gamma_{S}\, .
\end{equation}

As regards possible experimental implementations of the proposed setup and its
parameter regime, we would
like to mention that, typically, quantum dots are
made out of semiconducting heterostructures,
which satisfy above inequalities\cite{Kouwenhoven}.
Furthermore, in recent experiments, it has been shown
that the fabrication of hybrid structures with semiconductor and
superconductor being tunnel-coupled is
possible\cite{{Takayanagi,Franceschi}}.
Other candidate
materials are e.g. carbon nanotubes which also show Coulomb blockade behavior
with $U$ and $\delta\epsilon$ being in the regime of interest here\cite{Dekker}.  The
present work might provide further motivation to implement the  structures
proposed here.

Our goal in the following is to calculate the stationary charge
current of pairwise spin-entangled electrons for two competing
transport channels, first for the desired transport of two
entangled electrons into different leads ($I_1$) and second for
the unwanted transport of both electrons into the same lead
($I_2$). We compare then the two competing processes and show how
their  ratio, $I_1/I_2$, depends on the various system parameters
and how it can be made large. An important finding is that when
tunneling of two electrons into different leads occurs, the
current is suppressed due to the  fact that  tunneling into the
dots will typically take place from different points ${\bf r}_{1}$
and ${\bf r}_{2}$ on the superconductor (see Fig.\ 1) due to the
spatial separation of the dots $D_{1}$ and $D_{2}$. We show that
the distance of separation $\delta r=|{\bf r}_{1}-{\bf r}_{2}|$
leads to an exponential suppression of the current via different
dots if $\delta r>\xi$ (see (\ref{I_{1}})), where
$\xi=v_{F}/\pi\Delta$ is the coherence length of a Cooper pair. In
the relevant regime, $\delta r<\xi$, however, the suppression is
only polynomial in the parameter $k_{F}\delta r$, with $k_{F}$
being the Fermi wavevector in the superconductor, and depends
sensitively on the dimension of the superconductor. We
find (see Section \ref{efficiency}) that the suppression is less
dramatic in lower dimensional superconductors where we find
asymptotically smoother power law suppressions in  $k_{F}\delta
r$. On the other hand, tunneling via the same dot implies $\delta
r=0$, but suffers  a suppression due to $U$ and/or $\Delta$. The
suppression of this current is given by the small parameter
$(\gamma_{l}/U)^2$ in the case $U<\Delta$, or by $
(\gamma_{l}/\Delta)^2$, if $U>\Delta$ as will be derived in the
following. Thus, to maximize the efficiency of the entangler, we
also require $k_F\delta r<\Delta/\gamma_{l}, U/\gamma_{l}$.

Finally, we will discuss the effect of a magnetic flux on the
entangled current in an Aharonov-Bohm loop, and we will see that
this current contains both, single- and two-particle Aharonov-Bohm
periods whose amplitudes have different parameter dependences.
This allows us to distinguish processes where two electrons travel
through the same arm of the loop from the desired processes where
two electrons travel through different arms. The relative weight
of the amplitudes of the two Aharonov-Bohm periods are directly
accessible by flux-dependent current measurements which are then a
direct probe of the desired nonlocality of the entangled
electrons.

\section{Hamiltonian of the Andreev Entangler}
\label{Hamiltonian}

We use a tunneling Hamiltonian description of the system,
$H=H_{0}+H_{T}$, where
\begin{equation}
H_{0}=H_{S}+\sum_{l}H_{Dl}+\sum_{l}H_{Ll},\qquad l=1,2.
\end{equation}
Here, the superconductor is described by the
BCS-Hamiltonian\cite{Schrieffer}
$H_{S}=\sum_{{\bf k},\sigma}E_{\bf k}  \gamma_{{\bf k}\sigma}^{\dagger}
\gamma_{{\bf k}\sigma}$,
where $\sigma=\uparrow,\downarrow$, and the quasiparticle operators
$\gamma_{{\bf
k}\sigma}$ describe excitations out of the BCS-groundstate $|0\rangle_{S}$
defined by $\gamma_{{\bf k}\sigma}|0\rangle_{S}=0$. They are related to the
electron annihilation and creation operators $c_{{\bf k}\sigma}$ and
$c_{{\bf
k}\sigma}^{\dagger}$ through the Bogoliubov transformation \cite{Schrieffer}
\begin{eqnarray}
c_{{\bf k}\uparrow}&=&u_{\bf k}\gamma_{{\bf k}\uparrow}+v_{\bf
k}\gamma_{-{\bf k}\downarrow}^{\dagger}\nonumber\\
c_{-{\bf k}\downarrow}&=&u_{\bf k}\gamma_{-{\bf k}\downarrow}-
v_{\bf k}\gamma_{{\bf k}\uparrow}^{\dagger}\, ,
\label{Bogoliubov}
\end{eqnarray}
where $u_{\bf k}=(1/\sqrt{2})(1+\xi_{\bf k}/E_{\bf k})^{1/2}$ and
$ v_{\bf k}=(1/\sqrt{2})(1-\xi_{\bf k}/E_{\bf k})^{1/2}$ are the usual BCS
coherence factors \cite{Schrieffer}, and $\xi_{\bf k}=\epsilon_{\bf
k}-\mu_{S}$ is the normal state single-electron energy counted from the
Fermi level $\mu_{S}$, and $E_{\bf k}=\sqrt{\xi_{\bf k}^2+\Delta^2}$ is the
quasiparticle energy. We choose energies such that $\mu_{S}=0$. Both dots
are represented as one localized (spin-degenerate) level with energy
$\epsilon_{l}$ and is modeled by an Anderson-type Hamiltonian
$H_{Dl}=\epsilon_{l}\sum_{\sigma}d_{l\sigma}^{\dagger}d_{l\sigma}+Un
_{l\uparrow}n_{l\downarrow}$, $l=1,2$. The resonant dot level $\epsilon_{l}$
can be
tuned by the gate voltage. Other levels of the dots do not participate in
transport if $\delta\epsilon>\Delta\mu>k_{B}T$,
where $\Delta\mu=-\mu_{l}$, and $\mu_{l}$ is the chemical potential of lead
$l=1,2$,  and $\delta\epsilon$ is the single particle energy level spacing
of the
dots. The leads $l=1,2$ are assumed to be noninteracting (normal) Fermi
liquids with Hamiltonian
$H_{Ll}=\sum_{{\bf k}\sigma}\epsilon_{\bf k}a_{l{\bf
k}\sigma}^{\dagger}a_{l{\bf
k}\sigma}$. Tunneling from the dot $l$ to the lead $l$ or to the point ${\bf
r}_{l}$
in the superconductor is described by the tunnel
Hamiltonian $H_{T}=H_{SD}+H_{DL}$ with
\begin{eqnarray}
H_{SD}&=&
\sum\limits_{l\sigma} T_{SD}
d_{l\sigma}^{\dagger}\psi_{\sigma}({\bf r}_{l})+{\rm H.c.},
\label{tunnelhamilton1}\\
H_{DL}&=&
\sum\limits_{l{\bf k} \sigma} T_{DL} a_{l{\bf k}\sigma}^{\dagger}
d_{l\sigma}+{\rm H.c.}\, .
\label{tunnelhamilton2}
\end{eqnarray}
Here, $\psi_{\sigma}({\bf r}_{l})$ annihilates an electron with spin
$\sigma$ at
site ${\bf r}_{l}$,  and $d_{l\sigma}^{\dagger}$ creates it again (with the
same
spin) at dot $l$ with amplitude $T_{SD}$. $\psi_{\sigma}({\bf r}_{l})$ is
related to $c_{{\bf k}\sigma}$  by the Fourier transform $\psi_{\sigma}({\bf
r}_{l})=\sum_{\bf k} e^{i{\bf kr}_{l}}c_{{\bf k}\sigma}$.
Tunneling from the
dot to the state ${\bf k}$ in the lead is described by the tunneling  amplitude
$T_{DL}$.
We assume that the ${\bf k}$-dependence of  $T_{DL}$
can be safely neglected.
\section{Stationary Current and T-matrix }
\label{T-matrix}

The stationary current of {\it two}  electrons passing from the
superconductor
via virtual dot states to
the leads is given by
\begin{equation}
I=2e\sum\limits_{f,i}W_{fi} \rho_{i}\, ,
\label{current}
\end{equation}
where $W_{fi}$  is the transition rate from the superconductor to the
leads. We calculate this transition rate in a
T-matrix approach \cite{Merzbacher},
\begin{equation}
W_{fi}=2\pi\, |\langle
f|T(\varepsilon_i)|i\rangle|^2\,\delta(\varepsilon_f-\varepsilon_i)\,  .
\label{rate}
\end{equation}
Here,
$T(\varepsilon_i)=H_{T}\frac{1}{\varepsilon_i +i \eta-H}(\varepsilon_i-H_{0})$,
is the on-shell transmission or T-matrix,
with $\eta$ being a small positive real number which we take to zero
at the end of the calculation.
Finally,  $\rho_{i}$ is the stationary
occupation probability for the entire system to be in the state $|i\rangle$.
The T-matrix $T(\varepsilon_i)$ can be expanded in powers of the
tunnel Hamiltonian $H_{T}$,
\begin{equation}
\label{series}
T(\varepsilon_i)=H_{T}+H_{T}\sum_{n=1}^{\infty}
(\frac{1}{\varepsilon_i+i\eta-H_{0}}H_{T})^n \, ,
\end{equation}
where the initial energy is $\varepsilon_i=2\mu_{S}\equiv 0$. We work in the
regime defined in Eq. (\ref{regime}), i.e. $\gamma_{l}>\gamma_{S}$, and
$\Delta,U,
\delta\epsilon>\Delta\mu>\gamma_{l}, k_{B}T $,
and around the particular resonance $\epsilon_{l}\simeq \mu_{S}$.
Further, $\gamma_{S}=2\pi\nu_{S}|T_{SD}|^2$ and
$\gamma_{l}=2\pi\nu_{l}|T_{DL}|^2$
define the tunneling rates between superconductor and dots, and between dots
and
leads, respectively, with $\nu_{S}$ and $\nu_{l}$ being the DOS per spin at
the chemical potentials $\mu_{S}$ and $\mu_{l}$, respectively.
We will show that the total effective tunneling
rate  from the superconductor to the leads  is given by
$ \gamma_{S}^2/\gamma_{l}$
due to the Andreev process. To specify the initial state $|i\rangle$ of the system we point out that since $\Delta>\Delta\mu, k_{B}T$, the superconductor contains no quasiparticle initially. Also, due to the asymmetric barrier case, $|T_{DL}|>|T_{SD}|$ (or $\gamma_{l}>\gamma_{S}$), an electron on the dot level $\epsilon_{l}$ leaves the dot to the leads much faster than new electrons can be provided by the superconductor. In addition, a stationary occupation of the resonant levels $\epsilon_{l}$ is given by the grand canonical distribution function $\propto\exp(-\Delta\mu/k_{B}T)$ which is exponentially small if $\Delta\mu>k_{B}T$. This implies $\rho_{i}\simeq 0$ for initial states with occupied levels $\epsilon_{l}$. Therefore we consider the initial state
$|i\rangle=|0\rangle_{S}|0\rangle_{D}|\mu_{l}\rangle_{l}$,  where
$|0\rangle_{S}$ is the quasiparticle vacuum for the superconductor,
$|0\rangle_{D}$
means that both dot levels $\epsilon_{l}$ are unoccupied, and
$|\mu_{l}\rangle_{l}$
defines the occupation of the leads which are filled with electrons up to
the chemical potential $\mu_{l}$.
 We
remark that in our regime of interest no Kondo effects appear which could
destroy the spin entanglement, since
our dots contain each an even number of electrons in the stationary limit.


\section{Current due to tunneling to different leads}


We now calculate the current for the simultaneous tunneling of two
electrons into different leads. Since we assume that the spin is a
good quantum number we can specify the final state for two
electrons, one of them being in lead 1 the other in lead 2,
according to their total spin $S$. This spin can be either a
singlet (in standard notation)
$|S\rangle=(|\!\uparrow\downarrow\rangle - |\!
\downarrow\uparrow\rangle)/\sqrt{2}$ with $S=0$, or a triplet with
$S=1$. In the regime of interest (\ref{regime}), and since the total
spin is conserved, $[{\bf S}^2, H]=0$, the singlet state of the
initial Cooper pair will be conserved in the transport process and
the final state must also be a singlet. That this is indeed true
we can see explicitly if we allow for the possibility that the
final state could also be the $S_z=0$  triplet
$|t_{0}\rangle=(|\!\uparrow\downarrow\rangle +
|\!\downarrow\uparrow\rangle)/\sqrt{2}$. The not-entangled
triplets $|t_{+}\rangle=|\!\uparrow\uparrow\rangle$ and
$|t_{-}\rangle=|\!\downarrow\downarrow\rangle$ can be excluded
right away since the tunnel Hamiltonian $H_{T}$ conserves the
spin-component $\sigma$ and an Andreev process involves tunneling
of two electrons with different spin $\sigma$. Therefore we
consider two-particle final states of the form

\begin{equation}
|f\rangle=(1/\sqrt{2})\left(a_{1{\bf p}\uparrow}^{\dagger}a_{2
{\bf q}\downarrow}^{\dagger}\pm a_{1{\bf p}\downarrow}^{\dagger}a_{2{\bf
q}\uparrow}^{\dagger}\right)|i\rangle.
\end{equation}

The $-$ and $+$ signs belong to the
singlet $|S\rangle$ and triplet $|t_{0}\rangle$, resp. Note that this singlet/triplet
state is formed out of two electrons, one being in the  ${\bf p}$ state in lead 1
and with energy $\epsilon_{\bf p}$,
while the other one is in the ${\bf q}$ state in lead 2 with energy
$\epsilon_{\bf q}$. Thus, the two electrons are entangled in spin space
while separated in orbital space, thereby providing a non-local EPR pair.
The
tunnel process to different leads appears in the following order. A Cooper
pair breaks up due to tunneling of an electron with spin $\sigma$ to one
of the
dots (with empty level $\epsilon_{l}$) from the point of the superconductor
nearest to this dot. This is a virtual
state with energy deficit $ E_{\bf k}>\Delta$. Since $\Delta>\gamma_{l}$,
the second electron of the Cooper pair with spin $-\sigma$
tunnels to the other empty dot-level {\em before} the electron with spin $\sigma$ can
escape to the
lead. Therefore, both electrons tunnel almost simultaneously to the dots
(within the uncertainty time $\hbar/\Delta$).
Since we work at the resonance $\epsilon_{l}\simeq \mu_S= 0$ the energy
denominators in
(\ref{series}) show divergences $\propto 1/\eta$ indicating that tunneling
between
the dots and the leads is resonant and we have to treat
tunneling to all orders in
$H_{DL}$ in (\ref{series}). As a result $\eta$ will be replaced by
$\gamma_l/2$.
Tunneling back to the
superconductor is unlikely since $|T_{SD}|< |T_{DL}|$. We
can therefore write the transition amplitude between initial and final state
as
\begin{eqnarray}
\label{amplitude1}
\langle f|T_0|i\rangle&=&\frac{1}{\sqrt{2}}\langle i|a_{2{\bf
q}\downarrow}a_{1{\bf
p}\uparrow} T^{'} d_{1\uparrow}^{\dagger}d_{2\downarrow}^{\dagger}|i\rangle\nonumber\\
&&\times \langle i| (d_{2\downarrow}d_{1\uparrow}\pm
d_{2\uparrow}d_{1\downarrow}) T^{''}|i\rangle  \, ,
\end{eqnarray}
where $T_0=T(\varepsilon_i=0)$, and
the partial T-matrices $T^{'}$ and $T^{''}$ are given by
\begin{equation}
\label{T''}
T^{''}=\frac{1}{i\eta-H_{0}}H_{SD}\frac{1}{i\eta-H_{0}}H_{SD}\,  ,
\end{equation}
and
\begin{equation}
T^{'}=H_{DL}\sum\limits_{n=0}^{\infty} \left(\frac{1}
{i\eta-H_{0}}H_{DL}\right)^{2n+1}\, .
\end{equation}
In (\ref{amplitude1}) we used that the matrix element containing $T^{'}$ is
invariant under spin exchange $\uparrow\leftrightarrow\downarrow$. The part
containing $T^{''}$ describes the Andreev process of tunneling of two electrons with opposite spins from the SC to different dots 1,2, while the part containing
$T^{'}$ is the resonant dot $\leftrightarrow$  lead tunneling.

We first
consider the Andreev process. We insert a complete set of
single-quasiparticle (virtual) states
 between the two $H_{SD}$ in
(\ref{T''}) and find
for the singlet
 final state
\begin{eqnarray}
&&\langle i| (d_{2\downarrow}d_{1\uparrow}-
d_{2\uparrow}d_{1\downarrow}) T^{''}|i\rangle \nonumber\\
&&\qquad\qquad =
\frac{4T_{SD}^2}{\epsilon_{1}+\epsilon_{2}-i\eta}\sum\limits_{{\bf
k}} \frac{u_{\bf k}v_{\bf k}}{E_{\bf k}}
\cos{({\bf k}\cdot\delta {\bf r})}\, ,
\label{Andreev1}
\end{eqnarray}
where $\delta {\bf r}={\bf r}_{1}-{\bf r}_{2}$ denotes the distance
vector between the points on the superconductor from which
electron 1 and 2 tunnel into the dots. Note that the triplet contribution
vanishes since $u_{\bf k}v_{\bf k}=u_{-{\bf k}}v_{-{\bf k}}$ for s-wave superconductors.
The sum over ${\bf k}$ in (\ref{Andreev1}) can be calculated by linearizing the spectrum around the Fermi energy and we obtain

\begin{equation}
\label{Andreev2}
\sum\limits_{{\bf
k}} \frac{u_{\bf k}v_{\bf k}}{E_{\bf k}}
\cos{({\bf k}\cdot\delta {\bf r})} =
\frac{\pi}{2}\nu_{S}
\frac{\sin(k_{F}\delta r)}{k_{F}\delta r}
e^{-(\delta r/\pi\xi)}\, ,
\end{equation}
 where $k_{F}$ is the Fermi wavevector in the superconductor.
\section{Dominant  resonant tunneling events}

We turn to the calculation of the matrix
element in (\ref{amplitude1}) containing $T^{'}$ where tunneling is
treated to
all orders in $H_{DL}$. We introduce  the  ket notation $|1 2\rangle$, and, for simplicity, suppress the spin index $\sigma$. Here 1 stands  for quantum numbers
of the
electron on dot 1/lead 1 and similar for 2. For example, $|p q\rangle$ stands for
$a^\dagger_{1{\bf p}\sigma} a^\dagger_{2{\bf q}-\sigma}|i\rangle$, where
${\bf p}$ is from lead 1 and ${\bf q}$ from lead 2, or, correspondingly,  $|p D\rangle$
stands for $a^\dagger_{1{\bf p}\sigma} d^\dagger_{2,-\sigma}|i\rangle$, etc.
We restrict ourselves to the
resummation of
the following dot $\leftrightarrow$ lead transitions $|DD\rangle\rightarrow
|LD\rangle\rightarrow|DD\rangle$ or
$|DD\rangle\rightarrow|DL\rangle\rightarrow|DD\rangle$. In this sequence,
$|DD\rangle$ is the state with one electron on dot 1 and the other one on dot
2, and
$|LD\rangle$ denotes a state where one electron is in lead 1 and the other
one on dot 2. We thereby exclude tunneling sequences of the kind $|DD\rangle\rightarrow
|LD\rangle\rightarrow|LL\rangle\rightarrow|LD\rangle\rightarrow|DD\rangle$
or $|DD\rangle\rightarrow
|LD\rangle\rightarrow|LL\rangle\rightarrow|DL\rangle\rightarrow|DD\rangle$,
where both electrons are {\it virtually} simultaneously in the leads as well
as the creation of electron-hole pair excitations out of the Fermi sea. We
showed in Ref.\cite{RSL} that such contributions are suppressed in the
regime (\ref{regime}) considered here by the small parameter $\gamma_{l}/\Delta\mu$. It is quite clear that electron-hole pair excitations, which in principal could spoil the entanglement between the original partner electrons, are suppressed in our regime of interest. The following qualitative argument is in order (for a detailed calculation, see \cite{RSL}).
Suppose an electron initially on, say, dot 1, tunnels to the lead 1. Instead of hopping back to the dot, thereby completing the sequence $|DD\rangle\rightarrow|LD\rangle\rightarrow|DD\rangle$, it disappears in a final state out in the lead (with  energy $\epsilon_{\bf p}\sim \epsilon_{1}$). Now another electron from the Fermi sea ,with energy $\epsilon<-\Delta\mu$, hops on the empty dot.  The creation of such an electron-hole pair involves an energy deficit of at least $\Delta\mu$ (for $\epsilon_{1}=0$). Since this is a virtual state it can only exist during the uncertainty time $\sim 1/\Delta\mu$. The relaxation of this energy deficit requires the tunneling of this ``Fermi sea'' electron from the dot back to the lead. Since the tunneling from the dot to the lead takes a time on the scale of $1/\gamma_{1}$ this process is suppressed since $\Delta\mu>\gamma_{1}$. The dominant contributions are then resummed in the following sequence
\begin{eqnarray}
&&\langle {pq}|T^{'}|DD\rangle\hspace*{5cm} \nonumber\\
&&=\left\{
\begin{array}{r}
\langle pq|H_{D_{1}L_{1}}|Dq\rangle
\langle Dq|\sum\limits_{n=0}^{\infty}
(\frac{1}{i\eta-H_{0}}H_{D_{1}L_{1}})^{2n}| Dq\rangle\\
\times\langle Dq|\frac{1}{i\eta-H_{0}}H_{D_{2}L_{2}}|DD\rangle
\end{array}\right.
\nonumber \\
&&\quad
\left.\begin{array}{r}
+\langle pq|H_{D_{2}L_{2}}|pD\rangle
\langle pD|\sum\limits_{n=0}^{\infty}
(\frac{1}{i\eta-H_{0}}H_{D_{2}L_{2}})^{2n}|pD\rangle\\
\times\langle pD|\frac{1}{i\eta-H_{0}}H_{D_{1}L_{1}}|DD\rangle
\end{array}
\right\}\nonumber\\
&&\times\langle DD|\sum\limits_{m=0}^{\infty}
(\frac{1}{i\eta-H_{0}}H_{DL})^{2m}|DD\rangle .
\label{resummation1}
\end{eqnarray}
Since the sums for the transition
$|DD\rangle\rightarrow|DD\rangle$  via the sequences
$|DD\rangle\rightarrow|LD\rangle\rightarrow|DD\rangle$  and
$|DD\rangle\rightarrow|DL\rangle\rightarrow|DD\rangle$
 are independent, we can write
all summations in (\ref{resummation1}) as geometric series which allow for explicit resummations and we obtain
\begin{equation}
\label{BreitWigner1}
\langle {pq}|T^{'}|DD\rangle=\frac{-T_{DL}^2(\epsilon_{1}+\epsilon_{2}-i \eta)}
{(\epsilon_{1}+\epsilon_{\bf q}-i\gamma_{1}/2)
(\epsilon_{2}+\epsilon_{\bf p}-i\gamma_{2}/2)}\,.
\end{equation}
 Thus,
we see that the resummations in (\ref{resummation1}) cancel all divergences like the
$(\epsilon_{1}+\epsilon_{2}-i\eta)$ denominator appearing in
(\ref{Andreev1}), and that, as expected,
the resummation of divergent terms leads effectively to the replacement
$i\eta\rightarrow i\gamma_{l}/2$ which makes the limit $\epsilon_{l}\rightarrow
0$ well-behaved.
 In (\ref{BreitWigner1}) we neglected a small logarithmic correction to the (bare) resonant levels $\epsilon_{l}$ which is given as the real part of the self energy $\Sigma_{l}=|T_{DL}|^2\sum_{{\bf k}}(i\eta-\epsilon_l-\epsilon_{\bf k})^{-1}$ with $|{\rm Re}\Sigma_{l}|\sim\gamma_{l}\ln(\epsilon_{c}/\Delta\mu)<\Delta\mu$ and is therefore not important. The energy $\epsilon_{c}$ is the conduction band cut-off.
Making use of
Eqs.\ (\ref{current},\ref{rate}) and of Eqs. (\ref{Andreev1},\ref{Andreev2}) and
(\ref{BreitWigner1}),
we finally obtain for the  current (denoted by $I_{1}$)  where each of
the two entangled electrons tunnels into a {\it different} lead \cite{RSL}
\begin{equation}
\label{I_{1}}
I_{1}=\frac{e\gamma_{S}^2\gamma}{(\epsilon_1+\epsilon_2)^2+\gamma^2/4}
\left[\frac{\sin(k_{F}\delta r)}{k_{F}\delta r}\right]^2
\exp\left(-\frac{2\delta r}{\pi\xi}\right)\, ,
\end{equation}
where, $\gamma=\gamma_1 +\gamma_2$. We note that Eq. (\ref{I_{1}}) also holds
for the case with $\gamma_1\neq\gamma_2$. The current becomes exponentially
suppressed with increasing distance $\delta r$ between the tunneling points on
the superconductor, the scale given by the Cooper pair coherence length $\xi$.
This does not pose severe restrictions for conventional s-wave materials with
$\xi$ typically being on the order of micrometers. More severe is the restriction
that $k_{F}\delta r$ should not be too large compared to unity, especially if
$k_F^{-1}$ of the superconductor  assumes a typical value on the order of a few
Angstroms. Still, since the
suppression in $k_{F}\delta r$ is only power-law like there is  a sufficiently
large regime on the nanometer scale for
$\delta r$ where the current
$I_{1}$ can assume a finite measurable value. The power law suppression of the current in $1/k_{F}\delta r$ is very sensitive to the dimension of the SC and we suspect that the suppression will be softened by going over to lower dimensional superconductors. We will address this issue in Section \ref{efficiency}. The current (\ref{I_{1}}) has
a two-particle Breit-Wigner resonance form
which assumes it maximum
value when $\epsilon_1=-\epsilon_2$ (see also Fig.\ 3, and note that $\mu_{S}\equiv 0$),
\begin{equation}
\label{I_{1max}}
I_{1}=\frac{4e\gamma_{S}^2}{\gamma}
\left[\frac{\sin(k_{F}\delta r)}{k_{F}\delta r}\right]^2
\exp\left(-\frac{2\delta r}{\pi\xi}\right)\, .
\end{equation}
This resonance at $\epsilon_1=-\epsilon_2$ clearly shows that the current
is a correlated two-particle effect (even apart from any spin correlation) as we
should expect from the Andreev process involving the coherent tunneling of two
electrons.
Together with the single-particle resonances in Eq.
(\ref{BreitWigner1}) and by using energy conservation $\varepsilon_{i}=\varepsilon_{f}=0$, which implies $\epsilon_{\bf p}=-\epsilon_{\bf q}$, we thus see that
the current is
carried by correlated pairs of electrons whose lead energies satisfy
$|\epsilon_{\bf p}-\epsilon_{1}| \lesssim\gamma_{1}$ and $|\epsilon_{\bf q}-\epsilon_{2}| \lesssim\gamma_{2}$.

A particularly interesting case occurs when the energies of the dots, $\epsilon_1$
and  $\epsilon_2$, are both tuned to zero, i.e. $\epsilon_1=\epsilon_2=\mu_S=0$. We
stress that in this case the electron in lead 1 and its spin-entangled partner in
lead 2 possess exactly the {\it same} orbital energy. We have shown
previously\cite{BLS} that this degeneracy  of orbital energies
is a crucial requirement
for  noise measurements in which the singlets can be detected by an enhanced
noise in the current (bunching) due to a symmetrical orbital wavefunction of the singlet state, whereas uncorrelated electrons,  or,  more generally,
electrons in a triplet state, lead to a suppression of noise (antibunching). Note that not all triplets are entangled states. Only the triplet with $S_{z}=0$ is entangled. Measurement of noise enhancement is therefore a unique signature of entanglement \cite{BLS}.

We remark again that the current $I_{1}$ is carried by electrons
which are entangled in spin space and spatially separated in orbital space. In other
words, the stationary current $I_1$ is a current of non-local spin-based EPR pairs.
Finally, we note that due to the singlet character of the EPR pair we
do not know whether the electron in, say, lead 1 carries an up or a down spin,
this can be revealed only by a spin-measurement.
Of course, any measurement of the spin of one (or both) electrons
will immediately destroy the singlet state and thus the entanglement.
Such a spin measurement (spin read-out) can  be performed e.g. by making use of the
spin filtering effect of quantum dots\cite{RSL2}.
The singlet state will also be destroyed by spin-dependent scattering (but
not by Coulomb exchange interaction in the Fermi sea\cite{BLS}). However, it is
known experimentally that
electron spins in a semiconductor environment show unusually long dephasing times
approaching microseconds and can be transported phase coherently over distances
exceeding 100$\mu m$
\cite{{Kikkawa1,Kikkawa2,Awschalom,Fiederling,Ohno}}. This distance is
sufficiently long for experiments performed typically
on the length scale of quantum confined nanostructures\cite{Kouwenhoven}.

\section{current  due to tunneling via the same dot}

In this section we calculate  the current for tunneling of two electrons via the same dot and subsequently into the same lead. We show that  such processes are suppressed by a
factor $(\gamma_{l}/U)^2$ and/or $(\gamma_{l}/\Delta)^2$ compared to the process
discussed in the preceding section. But, in contrast to the previous case,
we do
not get a suppression resulting from the spatial separation of the Cooper
pair on the superconductor, since  here the two electrons tunnel from the
same point either from ${\bf r}_{1}$ or
${\bf r}_{2}$ (see Fig.\ 2). As before, a tunnel process
starts by
breaking up a Cooper pair followed by an Andreev process with
two possible tunneling sequences, see Fig.\ 4, panel
a). In a first step, one electron tunnels from the superconductor
to, say, dot 1, and in a second step the second electron also tunnels
to dot 1. Now
two electrons are simultaneously on the {\it same} dot which  costs additional  Coulomb repulsion
energy $U$, thus this  virtual state is suppressed by $1/U$. Finally, the two
electrons leave dot 1 and tunnel into lead 1.
There is an alternative competing process, see Fig.\ 4, panel  b), which avoids the double occupancy.
Here, one
electron tunnels to, say, dot 1, and then the same
electron tunnels  further into lead 1, leaving an excitation on the superconductor
which costs additional gap energy $\Delta$ (instead of $U$),
before finally the second electron
tunnels from the superconductor via dot 1 into lead 1.

\begin{figure}[h]
\centerline{\psfig{file=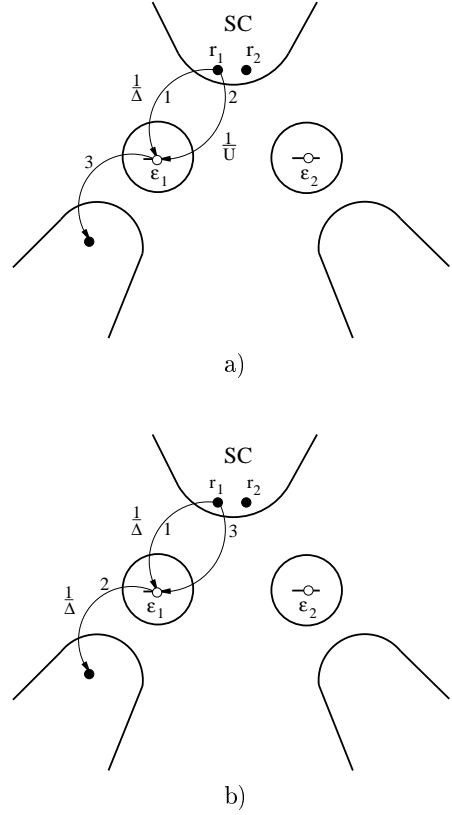,width=6cm}}
\vspace{0mm}
\caption{Two competing virtual processes are shown when the two electrons
tunnel via the same dot into the same lead:
a)  Andreev process
leading to a double occupancy of the dot with virtual energy suppression $ 1/U$,
and b)
the process which differs by the sequence of
tunneling, leading to an additional virtual energy suppression $ 1/\Delta$
instead of $1/U$.}
\end{figure}

We first concentrate on the  tunneling process b),
and note that the leading contribution comes from the processes
where both electrons have left the superconductor so that
the system has no energy deficit anymore. We still have to resum
the tunnel processes from the dot to the lead
to all orders in the tunnel Hamiltonian $H_{DL}$.
In what follows we suppress the label $l=1,2$ since the setup is assumed to be
symmetric and tunneling into either lead 1 or lead 2 gives the same result. The
transition amplitude $\langle f|T_0|i\rangle$ including only
leading terms is
\begin{eqnarray}
&&\langle f|T_0|i\rangle
=\nonumber\\
&&\sum\limits_{{\bf p}''\sigma}\langle
f|H_{DL}\sum\limits_{n=0}^{\infty}(\frac{1}{i\eta-H_{0}}H_{DL})^{2n}
|D{\bf p}''\sigma\rangle\nonumber\\
&&\times
\langle D{\bf p}''\sigma
|\frac{1}{i\eta-H_{0}}H_{SD}\frac{1}{i\eta-H_{0}}
H_{DL}\frac{1}{i\eta-H_{0}}H_{SD}|i\rangle\, ,
\label{series2}
\end{eqnarray}
where again $|f \rangle=(1/\sqrt{2})(a_{{\bf p}\uparrow}^{\dagger}a_{{\bf
p}'\downarrow}^{\dagger}\pm a_{{\bf p}\downarrow}^{\dagger}a_{{\bf
p}'\uparrow}^{\dagger})|i\rangle$, with $\pm$ denoting the triplet ($+$) and
singlet ($-$), resp., and the intermediate state $|D {\bf
p}''\sigma\rangle=d_{-\sigma}^{\dagger}a_{{\bf p}''\sigma}^{\dagger}|i
\rangle$. There are some
remarks in order regarding Eq. (\ref{series2}). The electron
which tunnels to the
state $|{\bf p}''\sigma\rangle$ has not to be resummed further since this would
lead either to a double occupancy of the dot which is suppressed by $1/U$, or
to the state with two
electrons simultaneously in the lead with a { \it virtual} summation over
the state ${\bf p''}$. But  we already mentioned above that the latter process
is suppressed by $\gamma_{l}/\Delta\mu$. The first factor in (\ref{series2}) describes therefore the multiple dot $\leftrightarrow$ lead tunneling of the electron with spin $-\sigma$ which resides on the dot in the intermediate state  $|D {\bf
p}''\sigma\rangle=d_{-\sigma}^{\dagger}a_{{\bf p}''\sigma}^{\dagger}|i
\rangle$ and which eventually tunnels in its final state in the lead. Again, this amplitude can be resummed explicitly with the result for $\sigma=\uparrow$,
\begin{eqnarray}
&&\langle f| H_{DL}\sum\limits_{n=0}^{\infty}
(\frac{1}{i\eta-H_{0}}H_{DL})^{2n}|D{\bf p}''\uparrow\rangle
\nonumber\\
&&\qquad\qquad
= -\frac{T_{DL}}{\sqrt{2}}
\frac{\epsilon_{l}+\epsilon_{{\bf
p}''}-i\eta}{\epsilon_{l}+\epsilon_{{\bf
p}''}-i\gamma_l/2}
\left(\delta_{{\bf p}''{\bf p}}\mp\delta_{{\bf p}''{\bf p}'}\right)\, ,
\label{uparrow}
\end{eqnarray}
and for $\sigma=\downarrow$ we get
\begin{eqnarray}
&&\langle f| H_{DL}\sum\limits_{n=0}^{\infty}
(\frac{1}{i\eta-H_{0}}H_{DL})^{2n}|D{\bf p}''\downarrow\rangle
\nonumber\\
&&\qquad\qquad
=\frac{T_{DL}}{\sqrt{2}}
\frac{\epsilon_{l}+\epsilon_{{\bf
p}''}-i\eta}{\epsilon_{l}+\epsilon_{{\bf
p}''}-i\gamma_l/2}
\left(\delta_{{\bf p}''{\bf p}'}\mp\delta_{{\bf p}''{\bf p}}\right).
\label{downarrow}
\end{eqnarray}
In (\ref{uparrow}) and (\ref{downarrow}) the upper sign belongs again
to the triplet and the lower sign to the singlet. For the amplitude containing the superconductor-dot transitions in (\ref{series2}) we obtain
\begin{eqnarray}
&& \langle D{\bf
p}''\sigma|\frac{1}{i\eta-H_{0}}H_{SD}\frac{1}{i\eta-H_{0}}H_
{DL}\frac{1}{i\eta-H_{0}}H_{SD}|i\rangle
\nonumber\\
&&
 =\frac{s\,T_{DL}T_{SD}^2\nu_{S}}{\Delta(\epsilon_{l}+\epsilon_{{\bf
p}''}-i\eta)},
\label{downarrow2}
\end{eqnarray}
where $s=+1$ (-1) for $\sigma=\uparrow$ $(\downarrow)$.
Combining the results  (\ref{uparrow})-(\ref{downarrow2}) we obtain for
the amplitude (\ref{series2})
\begin{equation}
\langle
f|T_0|i\rangle=
-\frac{2^{3/2}\nu_{S}(T_{SD}T_{DL})^2
(\epsilon_{l}-i\gamma_{l}/2)}
{\Delta(\epsilon_{l}+\epsilon_{\bf p}-i\gamma_{l}/2)
(\epsilon_{l}+\epsilon_{{\bf p}'}-i\gamma_{l}/2)}
\label{delta}
\end{equation}
for the singlet final state, whereas we get again zero for the
triplet.

Next we consider the process where the tunneling involves a double occupancy
of the dot, see panel a) in Fig.\ 4. In this case the transition
amplitude is
\begin{eqnarray}
&&\langle f|T_0|i\rangle=\nonumber\\
&&\sum\limits_{{\bf p}''\sigma}\langle f|H_{DL}\sum\limits_{n=0}^{\infty}
(\frac{1}{i\eta-H_{0}}H_{DL})^{2n}|D{\bf
p}''\sigma\rangle\nonumber\\
&&\times\langle D{\bf p}''\sigma|\frac{1}{i\eta-H_{0}}H_{DL}
\frac{1}{i\eta-H_{0}}H_{SD}\frac{1}{i\eta-H_{0}}H_{SD}|i\rangle\,.\nonumber\\
&&
\label{series3}
\end{eqnarray}
Repeating a similar calculation as before we find that the amplitude
is given by (\ref{delta}) but with $\Delta$ being replaced by $U/\pi$, and again, $\langle
f|T_0|i\rangle$ is only nonzero for the singlet final state.
We note that the two amplitudes (\ref{delta}) and (\ref{series3})
have the same initial and same final states and therefore have to be added coherently to obtain the total current
due to processes a) and b).
Then, using Eq.\ (\ref{current}) we find for the total current $I_2$
in case of tunneling of two electrons into the same lead \cite{correction},
\begin{equation}
I_{2}=\frac{e\gamma_S^2\gamma}{{\cal E}^2},\qquad
\frac{1}{\cal E}=\frac{1}{\pi\Delta}+\frac{1}{U}\, .
\label{I2}
\end{equation}
We see that the effect of the quantum dots shows up in the suppression
factor $(\gamma/{\cal E})^2$ for tunneling into the {\it same} lead.
We remark that in contrast to the previous case (tunneling into different
leads) the current
does not have a resonant behavior since the virtual dot states are no longer at
resonance due the energy costs $U$ or $\Delta$ in the tunneling process.
We now compare $I_{1}$ given in (\ref{I_{1max}}) with $I_{2}$ by forming the
ratio  of the two currents
\begin{equation}
\label{final}
\frac{I_{1}}{I_{2}}=
\frac{4{\cal E}^2}{\gamma^2}
\left[\frac{\sin(k_{F}\delta r)}{k_{F}\delta r}\right]^2   \exp\left(-\frac{2\delta
r}{\pi\xi}\right)\, .
\end{equation}
From this ratio we see that the desired regime  with $I_1$ dominating
$I_2$ is obtained when ${\cal E}/\gamma > k_F\delta r$, and $\delta r<\xi$.
We would like to emphasize that the relative
suppression of $I_2$ (as well as the absolute value of the current $I_1$) is
maximized by working around the resonances $\epsilon_{l}\simeq \mu_S= 0$\cite{incoherent} and, in addition, the desired injection of the two electrons at the same orbital energy is then achieved.

\section{Efficiency and Discussion}
\label{efficiency}

The current $I_{1}$ and therefore the ratio (\ref{final}) suffers an exponential suppression on the scale of $\xi$ if the tunneling of the two (coherent) electrons takes place from different points ${\bf r}_{1}$ and ${\bf r}_{2}$ of the superconductor. For conventional s-wave superconductors  the coherence length $\xi$ is typically on the order of micrometers and therefore poses not severe restrictions. So in the interesting regime the suppression of the Andreev amplitude is only polynomial $\propto 1/k_{F}\delta r$.  It was shown \cite{VanWees} that a superconductor on top of a two-dimensional electron gas (2DEG) can induce superconductivity (by the proximity effect) in the 2DEG with a finite order parameter. The 2DEG then becomes a two-dimensional (2D) superconductor. One could then desire to implement the two quantum dots in the 2DEG directly. More recently, it was suggested that superconductivity should also be present in ropes of single-walled carbon nanotubes \cite{Bouchiat} which are strictly one-dimensional (1D) systems. It is therefore interesting to calculate (\ref{Andreev2}) also in 2D and 1D. In the case of a 2D superconductor we evaluate (\ref{Andreev2}) in leading order in $\delta r/\pi\xi$ and find
\begin{eqnarray}
&&\sum\limits_{{\bf
k}({\tiny 2D})} \frac{u_{\bf k}v_{\bf k}}{E_{\bf k}}\cos{({\bf k}\cdot\delta {\bf r})}\nonumber\\
&&=\frac{\pi}{2}\nu_{S}\left(J_{0}(k_{F}\delta r)+2\sum\limits_{\nu=1}^{\infty}\,\frac{J_{2\nu}(k_{F}\delta r)}{\pi\nu}\right).
\label{Andreev3}
\end{eqnarray}

The right-hand side of (\ref{Andreev3}) can be approximated by $(\pi/2)\nu_{S}J_{0}(k_{F}\delta r)(1-(2/\pi)\ln 2)$ in the limit of large $k_{F}\delta r$. For large $k_{F}\delta r$, the behavior of the zeroth-order Besselfunction is $J_{0}(k_{F}\delta r)\sim\sqrt{2/\pi k_{F}\delta r}\,\cos(k_{F}\delta r-(\pi/4))$. So the amplitude decays asymptotically only $\propto 1/\sqrt{k_{F}\delta r}$ or the current $I_{1}$ by a factor $\propto 1/k_{F}\delta r$, respectively.
In the case of 1D we obtain

\begin{equation}
\sum\limits_{{\bf
k}(1D)} \frac{u_{\bf k}v_{\bf k}}{E_{\bf k}}\cos{({\bf k}\cdot\delta {\bf r})}=
\frac{\pi}{2}\nu_{S}\cos(k_{F}\delta r)\,e^{-(\delta r/\pi\xi)},
\end{equation}
where there are only oscillations and no decay of the Andreev amplitude (for $\delta r/\pi \xi<1)$.
We see that the suppression due to the finite separation of the tunneling points on the superconductor can be reduced considerably (or even excluded completely) by going over to lower-dimensional superconductors. By taking into account the dependence on the dimension of the superconductor we can relax the condition for the entangler to be efficient to
\begin{equation}
({\cal E}/\gamma)^{2}>(k_{F}\delta r)^{d-1},
\end{equation}
where $d$ is the dimension of the superconductor.\\
We note that the coherent injection of the two spin-entangled electrons by an Andreev process
via the dots into the leads allows for a time resolved detection of individual Cooper pairs in the leads since the delay time between the two partner electrons of a Cooper pair is given by $1/\Delta$ whereas the separation in time of subsequent Cooper pairs is given approximately by $2e/I_{1}\sim\gamma_{l}/\gamma_{S}^{2}$. Since $\Delta>\gamma_{S}$ and, in addition, $\gamma_{l}>\gamma_{S}$ the time delay between the two partners of the original Cooper pair is much shorter than the time difference between different Cooper pairs.


\section{Aharonov-Bohm oscillations }
In this section we show that the different tunneling paths of the
two electrons from the superconductor to the leads can be detected
via the flux-dependent Aharonov Bohm oscillations in the current
flowing through a closed loop (see Fig.\ 5). We show that due to
the possibility that two electrons can tunnel either via different
dots into different leads (non-local process) or via the same dot
into the same lead (local process), the current as a function of
magnetic flux $\phi$ penetrating the loop contains $h/e$ and
$h/2e$ oscillation periods. To be concrete, we consider a setup
where the two leads 1 and 2 are connected  such that they form an
Aharonov-Bohm loop, (see Fig.\ 5), where the electrons are injected
from the left via the superconductor, traversing the upper (lead
1) and lower (lead 2) arm of the loop before they rejoin to
interfere and then exit into the same lead, where the current is
then measured as a function of varying magnetic flux $\phi$.
\begin{figure}[h]
\centerline{\epsfig{file=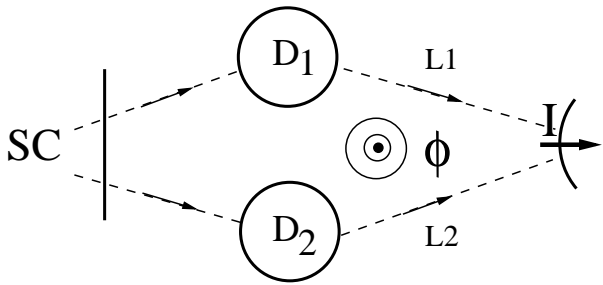,width=8cm}} \vspace{5mm}
\label{AB} \caption{The setup where the two outgoing leads 1 (L1) and 2 (L2) are connected to a common lead so that the tunneling path of the
electrons can form a loop. By applying a magnetic flux $\phi$ the
current shows Aharonov-Bohm oscillations with periods $h/e$ and
$h/2e$ which can be used to identify different tunneling paths of
the two electrons.}
\end{figure}
In the presence of a magnetic flux,  each tunneling amplitude
obtains a phase factor, $T_{D_1L_1}\rightarrow  T_{D_1L_1} e^{ i\phi/2\phi_0}$,
and $T_{D_2L_2}\rightarrow  T_{D_2L_2} e^{ -i\phi/2\phi_0}$,
where $\phi_0=h/e$ is the single-electron flux quantum.
For simplicity of the discussion we assume that the entire
phase is acquired  when the electron hops from the dot into the leads, so that
the process dot-lead-dot gives basically the full Aharonov-Bohm
phase factor $e^{\pm i\phi/\phi_0}$ of the loop (and only a negligible  amount of
phase is picked up
along the path from the superconductor to the dots). We stress that there is no loss of generality in this assumption.
The transition amplitude from the initial state to the final state has now the  following structure
$\langle f|T_0|i\rangle \sim
T_{D_1L_1} T_{D_2L_2}+T_{D_1L_1}^2e^{ i\phi/\phi_0} +
T_{D_2L_2}^2e^{ -i\phi/\phi_0}$.
Here, the first term comes from the process via different leads (see
(\ref{BreitWigner1})), where no Aharonov-Bohm phase is picked up.
The Aharonov-Bohm phase appears in the
remaining two terms, which come from processes via the same leads, either via lead
1 or lead 2 (see (\ref{delta}) and (\ref{series3})).
The total current $I$ is now obtained from $|\langle f|T_0|i\rangle|^2$ together with a summation over the final states, giving
$I=I_1 +I_2 +I_{AB}$, and the
flux-dependent Aharonov-Bohm current $I_{AB}$
is given by \cite{RSL}
\begin{eqnarray}
I_{AB}=   \sqrt {8I_1 I_2}F(\epsilon_l)
\cos {(\phi/\phi_0)} +  I_2 \cos {(2\phi/\phi_0)},
&&
\label{ABcurrent}\\
F(\epsilon_l)=\frac{\epsilon_l}{\sqrt{\epsilon_l^2+(\gamma_L/2)^2}},
&&
\label{F-factor}
\end{eqnarray}
where, for simplicity, we have assumed that
$\epsilon_1=\epsilon_2=\epsilon_l$, and
$\gamma_1=\gamma_2=\gamma_L$. Here, the first term (different
leads) is periodic in $\phi_0$ like for  single-electron
Aharonov-Bohm interference effects, while the second one (same
leads) is periodic in {\it half} the flux quantum $\phi_0/2$,
describing thus the interference of two coherent electrons
travelling the upper or the lower arm of the loop (similar single-
and two-particle Aharonov-Bohm effects occur in the Josephson
current through an Aharonov-Bohm loop\cite{CBL,CBL2}). It is clear
from (\ref{ABcurrent}) that the $h/e$ oscillation comes from the
interference between a contribution where the two electrons travel
through different arms with  contributions where the two electrons
travel through the same arm. Both Aharonov-Bohm oscillations with
period $h/e$, and $h/2e$, vanish with decreasing $I_2$, i.e. with
increasing on-site repulsion $U$ and/or gap $\Delta$. However,
their relative weight is given by $\sqrt{I_1/I_2}$, implying that
the $h/2e$ oscillations vanish faster than the $h/e$ ones. This
behavior is quite remarkable since it opens up the possibility to
tune down the unwanted leakage process $\sim I_2 \cos
{(2\phi/\phi_0)}$ where two electrons proceed via the same
dot/lead by  increasing $U$ with a gate voltage applied to the
dots. The dominant current contribution with period $h/e$ comes
then from the desired entangled electrons proceeding via different
leads. On the other hand, if $\sqrt{I_1/I_2}<1$, which could
become the case e.g. for $k_F\delta r>{\cal E}/\gamma$, we are
left with  $h/2e$ oscillations only. Besides the fact that the
Aharonov-Bohm oscillations are interesting in its own right, the
Aharonov-Bohm oscillations further provide an experimental probe
of the non-locality of the two spin-entangled electrons. Note that
dephasing processes which affect the orbital part suppress
$I_{AB}$. Still, the flux-independent current $I_1 + I_2$ can
remain finite and  contain electrons which are entangled in
spin-space, provided that there is only negligible spin-orbit
coupling so that the spin is still a good quantum number.

We would like to mention another important feature of the Aharonov-Bohm
effect under discussion, namely the relative phase shift between the
amplitudes of tunneling to the same lead and to different leads, resulting in the
additional prefactor $F(\epsilon_l)$ in the first term of the right-hand side of Eq.\
(\ref{ABcurrent}). This phase shift is due to the fact that there is a two-particle
resonance in the amplitude (\ref{BreitWigner1}) while there is only a single-particle
resonance in the amplitudes (\ref{delta}) and (\ref{series3}) (we recall that the second
resonance is suppressed by the Coulomb blockade effect). Thus, when the chemical potential
$\mu_{S}$ of the superconductor crosses the resonance,
$|\epsilon_l|\lesssim\gamma_L$, the amplitude
(\ref{BreitWigner1}) acquires an extra phase factor $e^{i\phi_r}$,
where $\phi_r=\arg[1/(\epsilon_l-i\gamma_L/2)]$. Then the interference
of the two amplitudes leads to the prefactor $F(\epsilon_l)=\cos\phi_r$
in the first term on the right-hand side of (\ref{ABcurrent}). In particular,
exactly at the middle of the resonance, $\epsilon_l=0$, the phase shift
is $\phi_r=\pi/2$, and thus the $h/e$ oscillations vanish, since $F(0)=\cos(\pi/2)=0$.
Note however, that although $F=\pm 1$ away from the resonance
($|\epsilon_l|>\gamma_L$) the $h/e$ oscillations vanish again,
now because the current $I_1\sim e\gamma_S^2\gamma_L/\epsilon_l^2$ vanishes.
Thus the optimal regime for the observation of the Aharonov-Bohm effect
is $|\epsilon_l|\sim\gamma_L$.

Finally, the preceding discussion shows that  even if the spins of two
electrons are entangled
their associated charge current does not reveal this spin-correlation
in a simple Aharonov-Bohm interference experiment\cite{ABdoubledot}.
Only if we consider the current-current correlations (noise) in a beam splitter setup,
can we detect also this spin-correlation in the transport current via its charge
properties\cite{BLS}.

\section{Andreev Entangler with Luttinger Liquid Leads}
We discuss a further implementation of an entangler for electron
spins. We replace the  quantum dots and the noninteracting leads
by interacting one-dimensional wires described as Luttinger
liquids (LL) (see Fig.\ 6).
\begin{figure}[h]
\centerline{\psfig{file=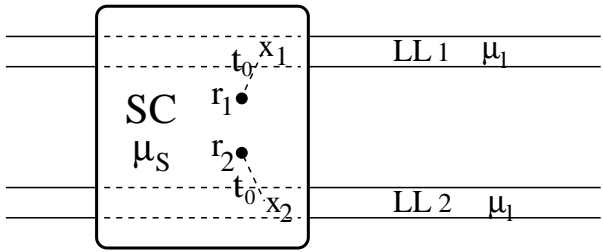,width=8cm}} \vspace{5mm}
\label{fig1} \caption{The entangler setup with Luttinger liquid
leads: An s-wave superconductor, SC, with chemical potential $\mu_{S}$
is weakly coupled to two strongly interacting bulk
Luttinger liquids 1,2 held at the same chemical potential
$\mu_{l}$. The two spin-entangled electrons of a Cooper pair can
coherently tunnel with amplitude $t_{0}$  by means of an Andreev
process from the superconductor to the leads. Two competing
transport channels are considered. Two electrons can tunnel to
different leads from points  ${\bf r}_{1}$ and ${\bf r}_{2}$ (with
distance $\delta r$) of the SC or from the same point
${\bf r}_{1}$ or ${\bf r}_{2}$  into the same lead. The
interaction between the two leads is assumed to be negligible.}
\end{figure}
The low energy excitations of these LL are collective charge and
spin density oscillations rather than quasiparticles which
resemble free electrons. As a consequence tunneling into a LL is
strongly suppressed due to interaction in the LL-leads.  We show
that the interaction can be used to separate two electrons,
originating from an Andreev process, so that they preferably
tunnel into different leads rather than into the same. Again we
take into account a finite tunneling distance on the
superconductor when the electrons tunnel to different leads.
\section{Model}
We consider an s-wave  superconductor as described in Section
\ref{Hamiltonian} with chemical potential $\mu_{S}$ which is
tunnel-coupled to two (spatially separated)  bulk LL-leads, both
held at the same chemical potential $\mu_{l}$ so that a finite
bias $\mu=\mu_{S}-\mu_{l}$ is applied. The Hamiltonian of the
whole system is  $H=H_{0}+H_{T}$, with
$H_{0}=H_{S}+\sum_{n=1,2}H_{Ln}$ describing the isolated
superconductor and LL-leads 1,2, resp.. Tunneling between the
superconductor and the leads is governed by the tunnel
Hamiltonian $H_{T}$, which will be described below. \\ The two
leads $1,2$ are assumed to be infinitely extended and interacting
one-dimensional systems described by conventional LL-theory. The
LL-Hamiltonian for the low energy excitations of lead $n=1,2$ is
written in a bosonized form as (neglecting backscattering)
\cite{Schulz}
\begin{eqnarray}
H_{Ln}-\mu_{l}N_{n}=\qquad\qquad&&\nonumber\\
\sum_{\nu=\rho,\sigma}\int\limits_{-L/2}^{+L/2}dx\,\left (\frac{\pi u_{\nu}K_{\nu}}{2}\,\Pi_{n\nu}^2+
\frac{u_{\nu}}{2\pi K_{\nu}}\,(\partial_{x}\phi_{n\nu})^2\right),&&
\label{LL Hamiltonian}
\end{eqnarray}
where $N_{n}$ is the number operator for lead $n=1,2$, and the
fields $\Pi_{n}(x)$ and $\phi_{n}(x)$ satisfy bosonic commutation
relations $\left [\phi_{n\nu}(x),\Pi_{m\mu}(x')\right
]=i\delta_{nm}\delta_{\nu\mu}\delta(x-x')$. The Hamiltonian
(\ref{LL Hamiltonian}) describes long-wavelength  charge
($\nu=\rho$) and spin ($\nu=\sigma$) density oscillations in the
LL propagating with velocities $u_{\rho}$ and $u_{\sigma}$, resp.
The velocities $u_{\nu}$ and the stiffness parameters $K_{\nu}$
depend on the interaction between the electrons in the LL.  In
the limit of vanishing backscattering, $u_{\sigma}=v_{F}$ and
$K_{\sigma}=1$ and the LL is described by only two parameters
$K_{\rho}<1$, and $u_{\rho}=v_{F}/K_{\rho}$, with $v_{F}$ being
the Fermi velocity. We decompose the field operator describing
electrons with spin $s$ into a right and left moving part,
$\psi_{ns}(x)=e^{ik_{F}x}\psi_{ns+}(x)+e^{-ik_{F}x}\psi_{ns-}(x)$.
The right (left) moving field operator $\psi_{ns+}(x)$
$(\psi_{ns-}(x))$  is then expressed as an exponential of bosonic
fields as \cite{Haldane,Heidenreich}

\begin{eqnarray}
\psi_{ns\pm}(x)&=\lim\limits_{\alpha\to
0}&\frac{\eta_{\pm,ns}}{\sqrt{2\pi\alpha}}
\exp\Big\{\pm\frac{i}{\sqrt{2}}\Big(\phi_{n\rho}(x)\Big.\Big.\nonumber\\
&&\Big.\Big.\quad+s\phi_{n\sigma}(x)\mp(\theta_{n\rho}(x)+s\theta_{n\sigma}(x))\Big)\Big\}\nonumber\\
&&
\label{bosonization}
\end{eqnarray}
where
$\partial_{x}\theta_{n\nu}(x)=\pi\Pi_{n\nu}(x)$. The
operators $\eta_{\pm,ns}$ are needed to ensure the correct
fermionic anticommutation relations. In (\ref{bosonization}) and hereafter we adopt the convention that $s=+1$ for $s=\uparrow$, and $s=-1$ for $s=\downarrow$, if $s$ has not the meaning of an index.\\ 
Transfer of electrons from
the SC to the LL-leads is described by the tunnel Hamiltonian
$H_{T}=t_{0}\,\sum_{ns}\,\psi_{ns}^{\dagger}\Psi_{s}({\bf r}_{n})
+ {\rm H.c.}$. The operator $\Psi_{s}({\bf r}_{n})$ annihilates an
electron with spin $s$ at the point ${\bf r}_{n}$ on the
superconductor nearest to the LL-lead $n$, and
$\psi_{ns}^{\dagger}$ creates it again with amplitude $t_{0}$ at
point $x_{n}$ in LL $n$ (see Fig.\ 6). We assume that the tunneling
amplitude $t_{0}$ does not depend on spin and also is the same for
both leads.\\

\section{Calculation of the Current:\\ two Competing Channels}

We now calculate the current for tunneling of two spin-entangled
electrons into different and into the same leads. Again, we use a
T-matrix approach introduced in Section \ref{T-matrix} and
calculate the current in lowest order in the tunneling Hamiltonian
$H_{T}$, which describes an Andreev process. For the current
$I_{1}$ for tunneling of two electrons into different LL-leads we
obtain in leading order in $\mu/\Delta$ and at zero temperature
\begin{equation}
I_{1}=\frac{I_{1}^{0}}{\Gamma(2\gamma_{\rho}+2)}\frac{v_{F}}{u_{\rho}}
\left[\frac{2\Lambda\mu}{u_{\rho}}\right]^{2\gamma_{\rho}},
\label{i_{1}}
\end{equation}
where $\Gamma$ is the Gamma function and $\Lambda$ is a short
distance cut-off on the order of the lattice spacing in the LL.
The interaction suppresses the current considerably and the bias
dependence has its characteristic non-linear form $I_{1}\propto
\mu^{2\gamma_{\rho}+1}$ with an interaction dependent exponent
$\gamma_{\rho}=(K_{\rho}+K_{\rho}^{-1})/4-1/2>0$. The parameter
$\gamma_{\rho}$ is the exponent for tunneling into the bulk of a
single LL, i.e.
$\rho(\varepsilon)\sim|\varepsilon|^{\gamma_{\rho}}$,  where
$\rho(\varepsilon)$ is the single particle DOS \cite{Schulz}. The
noninteracting limit $I_{1}^{0}$ is given as
\begin{equation}
I_{1}^{0}=\pi e\gamma^{2}\mu\,\left[\frac{\sin(k_{F}\delta r)}{k_{F}\delta r}\right]^2
\exp\left(-\frac{2\delta r}{\pi\xi}\right),
\end{equation}
with
 $\gamma=4\pi\nu_{S}\nu_{l}|t_{0}|^2$ being the probability per spin to
tunnel from the SC to the LL-leads and $\nu_{S}$ and $\nu_{l}$ are
the energy DOS per spin for the superconductor and the LL-leads at
the chemical potentials $\mu_{S}$ and $\mu_{l}$ resp.\,. We
again observe the same dependence on the tunneling distance $\delta r$
as before in the Andreev Entangler (see (\ref{I_{1}})). The
results for the lower dimensional superconductors are as before
and given in Section \ref{efficiency}.\\ Now we compare this
result with the process when the two electrons tunnel into the
same lead (with $\delta r=0$). Note that in this case the
correlation in time between subsequent tunneling events of the two
electrons of a Cooper pair becomes extremely important due to the
interaction in the LL-lead. We find after some calculation that
the current $I_{2}$ for tunneling into the same lead (1 or 2) is
suppressed if $\mu<\Delta$ with the result, in leading order in
$\mu/\Delta$,

\begin{equation}
I_{2}=I_{1}\sum\limits_{b=\pm 1}\,A_{b}\,\left(\frac{2\mu}{\Delta}\right)^{2\gamma_{\rho b}}.
\label{currentI222}
\end{equation}
The constant $A_{b}$ in (\ref{currentI222}) is of order one for not too strong interactions, but is decreasing with increasing interactions in the LL-leads, and is given by
\begin{eqnarray}
&&A_{b}=\frac{\pi}{2}\frac{\Gamma(2\gamma_{\rho}+2)}{\Gamma(2\gamma_{\rho
b} +2\gamma_{\rho}+1)}\left\{\left(\frac{\Gamma(\gamma_{\rho
b}+1)}{\Gamma(\frac{\gamma_{\rho
b}}{2}+1)\Gamma(\frac{1-\gamma_{\rho
b}}{2})}\right)^{2}\right.\nonumber\\
&&\left.+\sin^{2}\left(\frac{\gamma_{\rho
b}\pi}{2}\right)\Gamma^{2}(\gamma_{\rho
b}+1)\left(\frac{\Gamma\left(\frac{\gamma_{\rho
b}+1}{2}\right)}{\pi\Gamma\left(\frac{\gamma_{\rho
b}}{2}+1\right)}\right)^{2}\right\}.\nonumber\\ &&
\end{eqnarray}

We remark that in (\ref{currentI222}) the current $I_{1}$ is to be
taken at $\delta r=0$. The noninteracting limit
$I_{2}=I_{1}=I_{1}^{0}$ is rediscovered by putting
$\gamma_{\rho}=\gamma_{\rho b}=0$, and $u_{\rho}=v_{F}$. The
result for $I_{2}$ shows that the unwanted injection of two
electrons into the same lead is suppressed compared to $I_{1}$ by
a factor of $(2\mu/\Delta)^{2\gamma_{\rho +}}$, where
$\gamma_{\rho+}=\gamma_{\rho}$, if both electrons are injected
into the {\em same} branch (left or right movers), or by
$(2\mu/\Delta)^{2\gamma_{\rho -}}$ if the two electrons travel in
{\em different} directions.  Since it holds that $\gamma_{\rho
-}=\gamma_{\rho +}+(1-K_{\rho})/2>\gamma_{\rho +}$, it is more
favorable that the two electrons travel in the same direction
than that in opposite directions.  The suppression of the current
$I_{2}$ by $1/\Delta$ shows very nicely the two-particle
correlation effect in the LL, when the electrons tunnel into the
same lead. The larger $\Delta$, the shorter is the delay time
between the arrivals of the two partner electrons of a Cooper
pair, and, in turn, the more the second electron tunneling into
the same lead will feel the existence of the first one which is
already present in the LL. By increasing the bias $\mu$ the
electrons can tunnel faster through the barrier since there are
more channels available into which the electron can tunnel, and
therefore the effect of $\Delta$ is less pronounced. Also note
that this correlation effect disappears when interactions are
absent (i.e. when $\gamma_{\rho}=\gamma_{\rho b}=0$) in the LL.
Actual experimental systems which show LL-behavior are e.g.
metallic carbon nanotubes with similar exponents as derived here
\cite{Egger,Kane}. The exponents for tunneling into the bulk of a
LL and for tunneling into the end of a LL are, in general,
different, and it is predicted by theory \cite{Egger,Kane,Balents}
and consistent with experiment \cite{Nygard,Schoenenberger}, that
$\gamma_{\rho}^{bulk}<\gamma_{\rho}^{end}$. We therefore expect that
if the electrons tunnel into the end of the LL, the suppression is
even more pronounced, but to find the correct exponents would
require a careful recalculation in that geometry.\\ Finally, the
entangler setup is efficient if approximately
\begin{equation}
\left(\frac{\Delta}{2\mu}\right)^{2\gamma_{\rho +}}\,>\,\left(k_{F}\delta r\right)^{d-1},
\end{equation}
 where $d$ is the dimension of the superconductor, $k_{F}$ is the Fermi wavevector of
 the superconductor, and it is assumed that the coherence length
$\xi$ of the SC is large compared to $\delta r$ (see Section
\ref{efficiency}).

\section{Decay of the Electron-Singlet due to LL-interactions}
Here we consider the decay of a nonlocal singlet where one
electron is in lead 1 and the other in lead 2 due to the
interactions in the LL-leads. A quantity which accounts for this
consideration is the correlation function
\begin{equation}
P({\bf r},t)=\left|\langle S({\bf r},t)|S(0,0)\rangle\right|^{2}.
\end{equation}

Here, $P({\bf r},t)$ is the probability that a singlet state
injected at $t=0$ and at point ${\bf r}\equiv (x_{1},x_{2})=0$ is
found at some later time $t$ at point ${\bf r}$, and it is therefore a measure of how much of the initial singlet
state remains after switching on the interaction during the time
interval $t$. The singlet state created on top of the
LL-groundstates is
\begin{eqnarray}
|S({\bf r},t)\rangle&=&\sqrt{\pi\alpha}(\psi_{1\uparrow}^{\dagger}(x_{1},t)\psi_{2\downarrow}^{\dagger}(x_{2},t)\nonumber\\
&-&\psi_{1\downarrow}^{\dagger}(x_{1},t)\psi_{2\uparrow}^{\dagger}(x_{2},t))|0\rangle,
\label{singlet}
\end{eqnarray}
where the  extra normalization factor $\sqrt{2\pi\alpha}$ for the
singlet is introduced to guarantee $\int d{\bf r}\,P({\bf
r},t)=1$ in the noninteracting limit. The singlet-singlet correlation function factorizes into
a product of two single-particle correlation functions due to
negligible interactions between the two leads 1,2 and we obtain for $P({\bf r},t)$, retaining only nonoscillatory terms,
\begin{equation}
P({\bf r},t)=\prod_{n}\frac{1}{2}\sum\limits_{r=\pm}F(t)\,\delta(x_{n}-rv_{F}t),
\label{P}
\end{equation}
with a time decaying weight factor of the $\delta$-function
\begin{eqnarray}
F(t)&=&\prod_{\nu=\rho,\sigma}\sqrt{\frac{\Lambda^{2}}{\Lambda^{2}+(v_{F}-u_{\nu})^{2}t^{2}}}\nonumber\\
&\times&\left(\frac{\Lambda^{4}}{\left(\Lambda^{2}+
(v_{F}t)^{2}-(u_{\nu}t)^{2}\right)^{2}+\left(2\Lambda
u_{\nu}t\right)^{2}}\right)^{\gamma_{\nu}/2}.\nonumber\\ &&
\end{eqnarray}
Without interaction we have $F(t)=1$ which means that there is no
decay of the singlet state. As interactions  are turned on we see
that for times $t>\Lambda/u_{\rho}$ the singlet state starts to decay in
time. For long times $t$ and for $u_{\sigma}=v_{F}$, $K_{\sigma}=1$ and $u_{\rho}=v_{F}/K_{\rho}$, the asymptotic behavior of the decay is $F(t)\sim\frac{\Lambda}{u_{\rho}(1-K_{\rho})}[\frac{\Lambda^{2}}{u_{\rho}^{2}(1-K_{\rho}^{2})}]^{\gamma_{\rho}}[\frac{1}{t}]^{2\gamma_{\rho}+1}$, which for very strong interactions in the LL leads, i.e. $K_{\rho}$ much smaller than one, becomes $F(t)\sim[\Lambda/u_{\rho}t]^{2\gamma_{\rho}+1}$.
This result together with (\ref{P})  shows that charge and spin of
an electron propagate with velocity $v_{F}$, whereas charge (spin)
excitations of the LL propagate with $u_{\rho}$ ($u_{\sigma}$) as we confirm explicitly in the next Section. 

\section{Propagation of Charge and Spin}
Although the singlet state gets destroyed due to interaction the
spin information can still be transported through the wires by the
spin density excitations as we will show now. The normal ordered
charge density $\rho_{n}(x)$ of lead $n$ is
$\rho_{n}(x)=\sum_{sr}:\psi_{nsr}^{\dagger}(x)\psi_{nsr}(x):$,
where we kept  only the slow spatial variations of the density
operator, and $r=\pm$ denotes the branch of left- and
right-movers. Likewise, for the normal ordered spin density we
have
$\sigma_{n}^{z}(x)=\sum_{sr}s:\psi_{nsr}^{\dagger}(x)\psi_{nsr}(x):\,.$
We now consider a state
$|\Psi\rangle=\psi_{nsr}^{\dagger}(x_{n})|0\rangle$ where we
inject an electron with spin $s$ into branch $r$ of lead $n$, and
at time $t=0$, on top of the LL groundstate. We calculate the time
dependent charge and spin density fluctuations for this state and
find for the charge density fluctuations
\begin{eqnarray}
\label{chargeflukt}
&&(2\pi\alpha)\left\langle 0|\psi_{nsr}(x_{n})\rho_{n}(x_{n}',t)\psi_{nsr}^{\dagger}(x_{n})|0\right\rangle\nonumber\\
&&=\frac{1}{2}(1+rK_{\rho})\delta\left(x_{n}'-x_{n}-u_{\rho}t\right)\nonumber\\
&&+\frac{1}{2}(1-rK_{\rho})\delta\left(x_{n}'-x_{n}+u_{\rho}t\right),
\end{eqnarray}
and for the spin density fluctuations
\begin{eqnarray}
\label{spinflukt}
&&(2\pi\alpha)\left\langle 0|\psi_{nsr}(x_{n})\sigma_{n}^{z}(x_{n}',t)\psi_{nsr}^{\dagger}(x_{n})|0\right\rangle\nonumber\\
&&=\frac{s}{2}(1+rK_{\sigma})\delta\left(x_{n}'-x_{n}-u_{\sigma}t\right)\nonumber\\
&&+\frac{s}{2}(1-rK_{\sigma})\delta\left(x_{n}'-x_{n}+u_{\sigma}t\right).
\end{eqnarray}
We see that in contrast to the singlet, the charge and spin density
fluctuations in the LL created by the injected electron do not
decay and show a pulsed shape with no dispersion in time. This is
due to the linear energy dispersion relation of the LL-model. In
carbon nanotubes such a highly linear dispersion relation is
indeed realized and therefore carbon nanotubes should be well
suited for spin transport. Another interesting effect that shows
up in (\ref{chargeflukt}) and (\ref{spinflukt}) is the different
velocities of spin and charge, which is known as spin-charge
separation.  It would be interesting to test Bell inequalities
\cite{Bell} via spin-spin correlation measurements between the two
LL-leads and see if the initial entanglement of the spin singlet
is still observable in the spin density-fluctuations. Although
detection of single spins with magnitudes on the order of electron
spins has still not been achieved, magnetic resonance force
microscopy (MFRM) seems to be very promising in doing so
\cite{Rugar}. Another scenario is to use the LL just as an
intermediate medium which is needed to first separate the two
electrons of a Cooper pair and then to take them (in general other
electrons) out again into two (spatially separated) Fermi liquid
leads where the (possibly reduced)  spin entanglement could be
measured via the current noise in a beamsplitter experiment
\cite{BLS}. In this context we note that the decay of the singlet
state given by (\ref{P}) sets in almost immediately after the
injection into the LLs (the time scale is approximately the
inverse of the Fermi energy) but at least at zero temperature, the
suppression is only polynomial in time which suggests that some
fraction of the singlet state can still be recovered.

\section{Conclusion}

We discussed setups for creating  spin-entanglement
 in quantum confined nanostructures like
semiconductor quantum dots and mesoscopic wires based on
superconductors which provide a natural source of
spin-entanglement in form of Cooper pairs. We showed that a
superconductor can induce tunable nonlocal spin-entanglement
between two electron spins which reside on different quantum dots
without having a direct tunnel coupling between the dots. We then
described an entangler device that creates mobile spin-entangled
electrons via an Andreev process into different dots which are
tunnel-coupled to leads. The unwanted process of both electrons
tunneling into the same leads can be suppressed by increasing the
Coulomb repulsion on the quantum dots. We have shown that there
exists a regime of experimental relevance where the current of
entangled electrons shows a resonance and assumes a finite value
with both partners of the singlet being in different leads but
having the same orbital energy. This entangler thus satisfies the
requirements needed to detect spin entanglement via transport and
noise measurements. Further, we discussed the flux-dependent
oscillations  of the current in an Aharonov-Bohm loop which could
serve as an experimental means to detect the nonlocality of the
two correlated electrons. Finally, we discussed a novel setup
consisting of two interacting Luttinger liquid leads (such as
nanotubes) weakly coupled to a superconductor to separate the two
spin-entangled electrons. We found that the coherent subsequent
tunneling of two
 electrons into the same LL is suppressed if the
applied bias is smaller than  the superconducting gap in a
characteristic power-law manner.

Acknowledgment: We thank C. Bruder, W. Belzig, and E.V. Sukhorukov
for useful discussions. This work is supported by the Swiss NSF,
DARPA, and ARO.


\begin{references}
\bibitem{correction}
The result of $I_{2}$ is just half the value predicted earlier in \cite{RSL} due to a double counting of possible (orthogonal) final states therein.
\bibitem{incoherent}
We remark that incoherent
transport (sequential tunneling) is negligible as long as the scattering
rate inside the dots, $\Gamma_{\varphi}$, is much smaller than $\gamma_{l}$, since
$I_{{\rm seq}}/I_{{\rm coh}}\simeq \Gamma_{\varphi}/\gamma_{l}$
\cite{Datta}.
\bibitem{ABdoubledot}
We note, however, that the Aharonov-Bohm current can be used as a probe to detect
localized spin-singlets in coupled double-dots\cite{LS}.
\end{references}
\end{document}